\documentclass{article}
\usepackage{ijcai24}

\usepackage{times}
\usepackage{soul}
\usepackage{url}
\usepackage[hidelinks]{hyperref}
\usepackage[utf8]{inputenc}
\usepackage[small]{caption}
\usepackage{graphicx}
\usepackage{amsmath}
\usepackage{amsthm}
\usepackage{booktabs}
\usepackage{algorithm}
\usepackage{algorithmic}
\usepackage[switch]{lineno}
\usepackage{amsfonts}  
\usepackage{multirow}
\usepackage{bbold}
\usepackage{mathtools}

\usepackage{atbegshi} 

\AtBeginShipout{%
  \AtBeginShipoutUpperLeft{%
    \raisebox{-35pt}[0pt][0pt]{%
      \hspace*{0pt}\makebox[\paperwidth]{%
        \footnotesize Proceedings of the Thirty-Third International Joint Conference on Artificial Intelligence (IJCAI-24)%
      }%
    }%
  }%
}

\title{A Semi-supervised Molecular Learning Framework for Activity Cliff Estimation}
\author{
    Fang Wu
    \affiliations
   Stanford University   
    \emails
    fangwu97@stanford.edu
}
\begin{document}

\maketitle

\begin{abstract}
Machine learning (ML) enables accurate and fast molecular property predictions, which is of interest in drug discovery and material design. Their success is based on the principle of similarity at its heart, assuming that similar molecules exhibit close properties. However, activity cliffs challenge this principle, and their presence leads to a sharp decline in the performance of existing ML algorithms, particularly graph-based methods. To overcome this obstacle under a low-data scenario, we propose a novel semi-supervised learning (SSL) method dubbed SemiMol, which employs predictions on numerous unannotated data as pseudo-signals for subsequent training. Specifically, we introduce an additional instructor model to evaluate the accuracy and trustworthiness of proxy labels because existing pseudo-labeling approaches require probabilistic outputs to reveal the model's confidence and fail to be applied in regression tasks. Moreover, we design a self-adaptive curriculum learning algorithm to progressively move the target model toward hard samples at a controllable pace. Extensive experiments on 30 activity cliff datasets demonstrate that SemiMol significantly enhances graph-based ML architectures and outpasses state-of-the-art pretraining and SSL baselines. 
\end{abstract}

\section{Introduction}
The Similar-Structure, Similar-Property Principle (SSP)~\cite{bender2004molecular} has been a central premise in medicinal chemistry that similar molecules tend to exhibit similar properties, which can be physical (\emph{e.g.}, boiling points) or biological (\emph{e.g.}, activity). This fundamental assertion is validated by long experience suggesting that rules of thumb such as phenethylamines are likely to have activity in the central nervous system and that $\beta$-lactams frequently possess antibacterial activity~\cite{martin2002structurally}. Computational chemists have also exploited this premise in their analysis of the molecular diversity of compound libraries and the selection of compounds for high-throughput screening (HTS)~\cite{johnson1989molecular}. 
\begin{figure}[t]
    \centering
    \includegraphics[width=1.0\columnwidth]{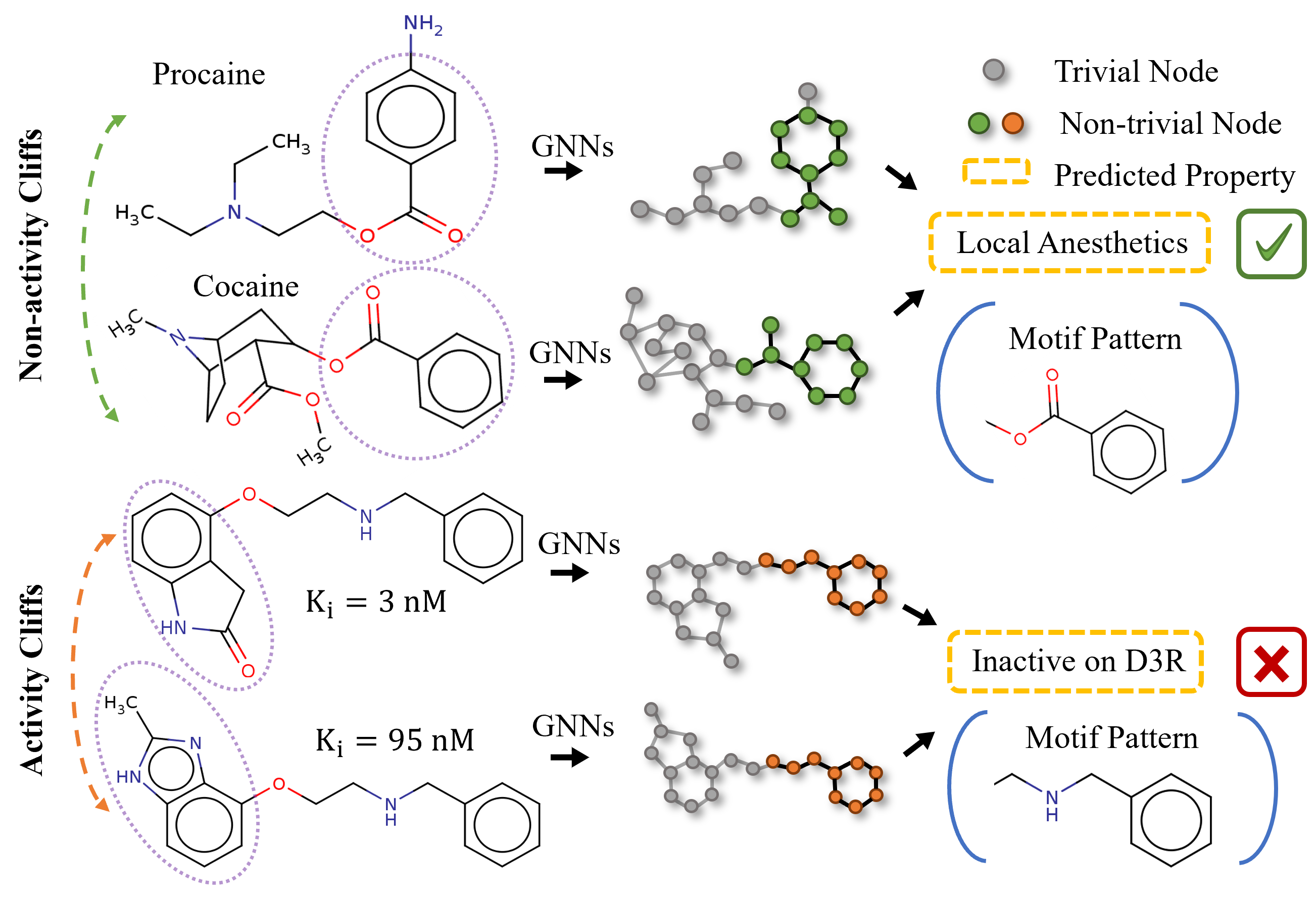} 
    \caption{For non-activity cliffs, GNNs successfully discover the motif of \texttt{COC(=O)c1ccccc1} and group molecules with this pattern as local anesthetics. For activity cliffs, GNNs are misguided by the shared motif pattern of \texttt{COCCNCc1ccccc1} and wrongly predict the inactivity of relevant molecules.}
    \label{fig:problem}
\end{figure}

This concept of chemical similarity (or molecular similarity) rationalizes the thriving deployment of machine learning (ML) models in discovering and designing new drugs with expected functions. Noticeably, ML algorithms usually work by recognizing the statistically important patterns and assigning close values to neighbor samples in the high-dimensional feature space~\cite{mitchell1997machine}. They learn motifs that appear repeatedly in the training set, cluster molecules that carry the same motif, and make similar predictions for them~\cite{wu2023molformer}. Therefore, SSP lays down a solid scientific foundation for and highly aligns with the inner operation logic of ML mechanisms.

Despite the triumph that ML has achieved in forecasting molecular properties, one particular exception to this principle holds great insights into the underlying structure-activity (or structure-property) relationships~\cite{stumpfe2020advances}. This exception is constituted by activity cliffs~\cite{maggiora2006outliers}, which are pairs of structurally similar molecules that exhibit a large difference in their biological activity. 
These outliers can impose a detrimental effect on ML models by misguiding them to seriously mispredict the activity of certain molecules, even with an overall high model predictivity (see Figure~\ref{fig:problem}). Although numerous studies have focused on defining activity cliffs and investigating their influence~\cite{van2022exposing}, how to ameliorate this obstacle to the development of ML remains unexplored.

In particular, the quantity and quality of the data are one of the most pivot factors in the success of deep learning (DL) algorithms, but the evaluation of DL methods on activity cliff estimation is always in a low-data scenario. Meanwhile, the graph research community for molecule modeling has been trying to replicate the victory of self-supervised pretraining in NLP and claims substantial improvements by pretraining on large-scale datasets~\cite{rong2020self,zhou2022uni}. Consequently, it provokes the question that \emph{``how much benefit self-supervised learning can bring to estimate activity cliffs?''}. As an answer, we observe a negligible reward gained by pretraining in many cases. Our first important finding is that self-supervised graph pretraining does not always have statistically significant advantages over non-pretraining methods with an average improvement ratio of merely 6.35\%. Secondly, although several recent studies leverage the 3D spatial information of molecules during the pretraining stages~\cite{fang2022geometry}, the incorporation of 3D geometry does not exhibit considerable advantages in recognizing molecular cliffs. Third, the design of the backbone GNNs has a larger impact on the accuracy of activity cliff estimation than the choice of pretraining mechanisms. 

This phenomenon inspires us to explore a more efficient schema to employ the abundant unlabeled molecules and relieve the data scarcity in activity cliff identification. Towards this direction, we present a novel and effective semi-supervised learning (SSL) mechanism dubbed SemiMol, which leverages predictions on unannotated data as supervised signals for subsequent training. However, pseudo-labels can be unreliable because labeled and unlabeled samples are usually drawn from different data distributions and DL models often struggle to detect this discrepancy. Besides that, existing pseudo-labeling approaches require probabilistic outputs to reveal the model's confidence and fail to be applied in regression tasks~\cite{rizve2021defense,wu2023instructbio}. To surmount this barrier, we introduce an additional instructor model to evaluate the accuracy and trustworthiness of proxy labels. These confidence scores then instruct the target molecular model to selectively focus on different data points in a progressive way. In SemiMol, rather than using fixed thresholds or manually tuning thresholds with percentile scores~\cite{cascante2021curriculum}, we propose a self-adaptive criterion for curriculum learning to determine which subset of pseudo-labeled samples should be incorporated into the training set. With these delicately designed controllable learning paces, the target molecular model can take advantage of the unlabeled data assuredly without suffering from over-reliance on labeled data and the severe inaccuracy of some pseudo-labels. 

We conducted extensive experimentation and visualization on MoleculeACE~\cite{van2022exposing} to verify the effectiveness of our method. It can be found that our SemiMol significantly promotes the representation ability of graph-based DL models in low-data scenarios and it outperforms state-of-the-art (SOTA) pretrained GNNs, descriptor-based ML models, and sequenced-based language models as well as prevailing SSL methods on all 30 activity cliff datasets. These results provide encouraging evidence that SemiMol can accurately capture adequate chemical and biological information to recognize cliffs in bioactivity. Our work positions itself in a broader movement within the graph-learning community for bioinformatics and aims to explore the optimal solution for low-data molecular property prediction systematically. 

\begin{figure*}
    \centering
    \includegraphics[width=0.85\textwidth]{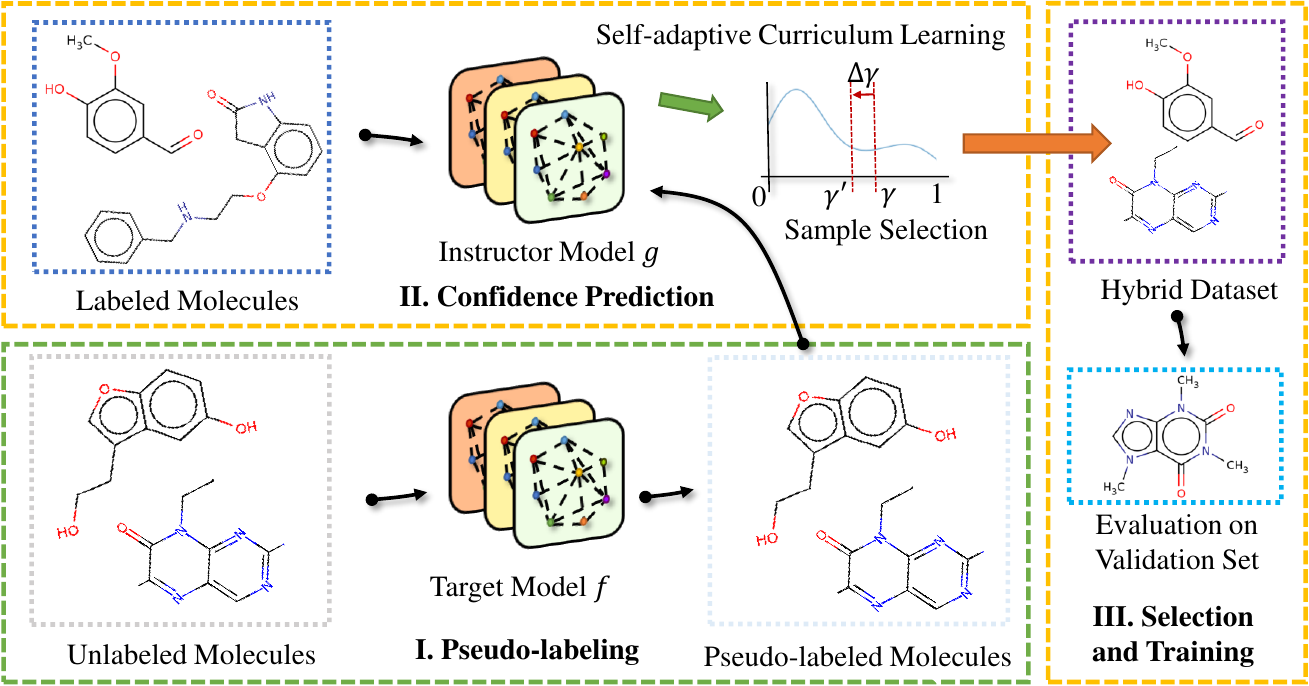}
    \caption{Illustration of SemiMol. The target model first assigns predictions for unlabeled molecular data. Then the instructor model analyzes the confidence of those proxy labels. After that, a self-adaptive curriculum learning schema is adopted to move the target model towards hard samples progressively. By iteratively repeating these processes, the target model can extract task-specific information from vast unannotated molecules to the greatest extent. }
    \label{fig:model}
\end{figure*}
\section{Method}
\subsection{Preliminaries and Background}
\paragraph{Task formulation.} Assume that there exists a set of accessible molecular data $\mathcal{D}^{\star} = \left\{\left(x_i^{\star}, y_i^{\star}\right)\right\}_{i=1}^N$, which consists of labeled data points $x_i^{\star}$. They can be in different formats such as 1D sequences, 2D graphs, and 3D structures, and $y_i^{\star}$ can be any discrete (\emph{e.g.}, toxicity, and drug reaction) or continuous properties (\emph{e.g.}, water solubility, and free energy). Here we only consider the continuous property to align with the setting of the activity cliff estimation problem. Still, our approach can be easily extended to binary or multi-label circumstances. The target molecular model $f:\mathcal{X}\rightarrow \mathcal{Y}$ aims to capture the mapping of molecules and their properties. It can be any type of architecture such as Transformers, GNNs, or geometric neural networks. We also have another set of unseen data $\mathcal{D}^{test}$ to evaluate the model's performance.
Typically, $\mathcal{D}^{\star}$ is divided into the training and validation sets, denoted as $\mathcal{D}^{train} = \left\{\left(x_i^{train}, y_i^{train}\right)\right\}_{i=1}^{N_1}$ and $\mathcal{D}^{val} = \left\{\left(x_i^{val}, y_i^{val}\right)\right\}_{i=1}^{N_2}$, respectively. Furthermore, there is a set of unlabeled data points $\mathcal{D}^{*} = \left\{x_i^{*}\right\}_{i=1}^{M}$, where the number of unlabeled data $M$ is much larger than that of labeled data $N$ (\emph{e.g.}, 100 or 10000 times larger). 

\paragraph{Semi-supervised learning.}  SSL has attracted increasing interest in overcoming the need for large annotated datasets. Existing semi-supervised algorithms can be roughly separated into three sorts: consistency regularization, proxy-label methods, and generative models. 
The consistency regularization is based on the simple concept that randomness within the neural network (\emph{e.g.}, with dropout) or data augmentation transformations should not modify model predictions given the same input and imposes an auxiliary loss. This line of research includes $\pi$-model~\cite{laine2016temporal}, temporal ensembling~\cite{laine2016temporal}, and mean teachers~\cite{tarvainen2017mean}
The proxy-label methods regard proxy labels on unlabeled data as targets and consist of two groups: self-training~\cite{yarowsky1995unsupervised}, where the model itself produces the proxy labels, and multiview learning~\cite{zhao2017multi}, where proxy labels are produced by models trained on different views of the data. 
The generative models rely on variational autoencoders (VAE)~\cite{ehsan2017infinite} and generative adversarial networks (GAN)~\cite{odena2016semi} to capture the joint distribution $p(y|x)$ more accurately. 

\subsection{Motivation of SemiMol} 
\paragraph{Proxy-labeling with confidence.} Labeled data $\mathcal{D}^{\star}$ and unlabeled data $\mathcal{D}^{*}$ in most cases follow significantly different data distributions, \emph{i.e.}, $\mathbb{P}(x_i^{\star}, y_i^{\star}) \neq \mathbb{P}(x_i^{*}, y_i^{*})$. However, on the one hand, many existing semi-supervised algorithms~\cite{yarowsky1995unsupervised,laine2016temporal,sohn2020fixmatch,xie2020self} assume that there is no distributional shift between $\mathcal{D}^{\star}$ and $\mathcal{D}^{*}$. They are likely only to reinforce the consistent information in the labeled data $\mathcal{D}^{\star}$ to unlabeled examples $\mathcal{D}^{*}$ instead of mining auxiliary information from $\mathcal{D}^{*}$, without exception for proxy labeling~\cite{lee2013pseudo}. On the other hand, despite the versatility and modality-agnostic of proxy labeling, it achieves relatively poor performance compared to recent semi-supervised algorithms~\cite{rizve2021defense}. This arises because some pseudo-labels $\hat{y}_i^{*}$ can be severely incorrect during training due to the poor generalization ability of classic DL models~\cite{hendrycks2021many}. If we directly utilize pseudo-labels that are predicted by a previously learned model for subsequent training, the conformance-biased information in the proceeding epochs could increase confidence in erroneous predictions and eventually lead to a vicious circle of error accumulation~\cite{arazo2020pseudo}. The situation can be even worse when the labeled data $\mathcal{D}^{\star}$ contain noises due to unavoidable experimental errors. Consequently, it is necessary to understand the quality and reliability of pseudo-annotations and intelligently select a subset to diminish the noise present in training~\cite{rizve2021defense}.   

\paragraph{Confidence for regression tasks.} Even though confidence is crucial for pseudo-label selection and confidence-based selection reduces pseudo-label error rates, the poor calibration of neural networks renders this solution insufficient. Explicitly, incorrect predictions in poorly calibrated networks can also have high confidence scores (\emph{i.e.}, $\hat{y}_i^{*}\rightarrow 0$ or $\hat{y}_i^{*}\rightarrow 1$)~\cite{rizve2021defense}. More importantly, the majority of existing approaches such as uncertainty-aware pseudo-label selection (UPS)~\cite{rizve2021defense} resort to the target model's output $\hat{y}_i$ as the indicator of confidence. They tend to produce hard labels by $y_i^{*} =\mathbb{1}\left[\hat{y}_i^{*} \geq \gamma_1 \right]$ or $y_i^{*} =\mathbb{1}\left[\hat{y}_i^{*} \leq \gamma_2 \right]$, where $\gamma_1$ and $\gamma_2$ are pre-defined confidence thresholds. However, this selection mechanism becomes inapplicable if $\mathcal{Y}$ is a continuous label space because networks no longer output class probabilities. In other words, for regression tasks, $\hat{y}_i^{*}$ does not disclose any confidential information. Noticeably, a great number of biochemical problems are regression-based, including molecular property prediction~\cite{wu2018moleculenet}, 3D structure prediction~\cite{jumper2021highly}, and binding affinity prediction~\cite{wang2005pdbbind}. This presents an unavoidable but essential challenge for probabilistic output-based proxy-labeling algorithms~\cite{rizve2021defense}. 

According to these two motivations, instead of depending on the output of the model $\hat{y}_i^{*}$ to judge the reliability of the proxy labels, we accompany the target molecular model $f$ with an additional instructor model $g$. It plays the role of a critic and predicts label observability, \emph{i.e.}, whether the label is true or fake. The introduction of $g$ disentangles the confidence prediction and the property prediction and can greatly reduce the noise introduced by the pseudo-labeling process.

\subsection{Framework of SemiMol}
\paragraph{Instructor-guided SSL.} To circumvent the obstacle of over-reliance and reduce the noise of proxy labels, we present an instructor-guided semi-supervised framework, dubbed SemiMol (see Figure~\ref{fig:model}), to intelligently select a subset of pseudo-labels with less noise. SemiMol is made up of two components: one target molecular model $f$ and one instructor model $g$. The former $f$ is responsible for predicting the properties $\hat{y}$. In contrast, the latter $g$ plays a role in measuring the reliability $p$ of supervised signals, which can be interpreted as the confidence probabilities of pseudo-labels. 
We separate the integral procedure of SemiMol into two phases. In the first step, we retain pseudo-labels $\left\{\hat{y}_i^{*}\right\}_{i=1}^{M}$. There are several approaches to creating proxy labels, such as label propagation via neighborhood graphs~\cite{iscen2019label}. Here we follow~\cite{lee2013pseudo} and require the molecular model $f$ to directly annotate samples in the unlabeled dataset $\mathcal{D}^{*}$. Then in the following step, we construct a new dataset with both labeled and pseudo-labeled samples as $\mathcal{D}'= \mathcal{D}^{\star} \cup \left\{\left(x_i^{*},\hat{y}_i^{*}\right)\right\}_{i=1}^{M}$ and proceed training the target molecular model $f$ and the instructor model $g$ based on this new set. These two operations are iteratively repeated until $f$ reaches the optimal performance on the validation set $\mathcal{D}^{val}$. 

To be specific, the instructor model $g:(\mathcal{X} \times \mathcal{Y} \times \mathcal{H}_f)\rightarrow \mathcal{P}$ forecasts the confidence measure $p_i$ ($0\leq p \leq 1$) of whether the given label $y'_i$ belongs to the ground-truth label set $\left\{y_i^{\star}\right\}_{i=1}^N$ or the pseudo-label set $\left\{y_i^{*}\right\}_{i=1}^{M}$. It digests three items: the data sample with its label $(x'_i, y'_i)\in \mathcal{D}'$ and an additional loss term $\mathcal{H}_f(f(x'_i), y'_i)$, where $\mathcal{H}_f$ is traditionally selected as a root-mean-squared-error (RMSE) loss or a mean absolute error (MAE) loss for regression tasks and cross-entropy (CE) loss for classification problems. Here we regard $\mathcal{H}_f(.)$ as the ingredient of $g$'s input to provide more information about the main molecular property prediction task. Finally, the instructor model $g$ is supervised via a binary CE loss (BCE) as
\begin{equation}
\label{equ:loss_instructor}
\begin{split}
    \mathcal{L}_g&\left(\mathcal{D}', \{\hat{y}_i'\}_{i=1}^{N+M}\right) = \sum_{(x'_i,y'_i)\in \mathcal{D}'}\textrm{BCE}(p_i, c_i) \\
    &= \sum_{(x'_i,y'_i)\in \mathcal{D}'}\textrm{BCE}\Big(g\big(x'_i, y'_i,  \mathcal{H}_f\left(\hat{y}_i', y'_i\right)\big), c_i\Big),
\end{split}
\end{equation}
where $c_i\in [0, 1]$ is an integer and represents the observability mask. It indicates whether $y_i$ is pseudo-labeled ($c_i=0$) or not ($c_i=1$). 

\begin{algorithm}[t]
  \caption{SemiMol Algorithm}
  \begin{algorithmic}[1]
    \STATE \textbf{Input:} target model $f$, instructor model $g$, labeled data $\mathcal{D}^{\star}$, unlabeled data $\mathcal{D}^{*}$, pseudo-label update frequency $k$, threshold percent $\gamma$, stepping threshold percent $\Delta \gamma$
    \STATE Initialize and pretrain a target model $f_0$ and an instructor model $g_0$
    \FOR{epochs $n=0,1,2,...$}
        \IF{$n\mod k == 0$}
            \STATE $\hat{y}_i^{*} \leftarrow f(x_i^*)\quad \forall x_i^* \in \mathcal{D}^{*}$  $\quad \vartriangleright$ Iteratively assign pseudo-labels every $k$ epochs
        \ENDIF
        \STATE $\mathcal{D}'\leftarrow \mathcal{D}^{\star} \cup \left\{\left(x_i^{*},\hat{y}_i^{*}\right)\right\}_{i=1}^{M}$ 
        \STATE $\hat{y}_i' \leftarrow  f(x'_i) \quad \forall x'_i \in \mathcal{D}'$
        \STATE $p_i \leftarrow g(x'_i, y'_i, \mathcal{H}_f(.) ) \quad \forall (x'_i, y'_i) \in \mathcal{D}'$ $\quad \vartriangleright$ Predict the confidence scores 
        \STATE $\mathcal{L}_g\left(\mathcal{D}', \{\hat{y}_i'\}_{i=1}^{N+M}\right) \leftarrow$ Equation~\ref{equ:loss_instructor}
        \STATE $\mathcal{D}'' \leftarrow \mathcal{D}^{\star} \cup \left\{\left(x_i^{*},\hat{y}_i^{*}\right) | p_i^{*} \geq \gamma \right\}$ $\quad \vartriangleright$ Build the training set 
        \STATE $\mathcal{L}_f\left(\mathcal{D}''\right) \leftarrow$ Equation~\ref{equ:loss}
        \STATE Update the parameters of $f$ and $g$ based on $\mathcal{L}_g(.)$ and $\mathcal{L}_f(.)$
        \STATE Evaluate $f$ on the validation set and attain performance $s$
        \IF{$n > 0$ \AND $s < s'$} 
            \STATE $\gamma \leftarrow \gamma - \Delta \gamma $ $\quad \vartriangleright$ Turn down the threshold automatically 
        \ENDIF
        \STATE $s' \leftarrow s$ $\quad \vartriangleright$ Record the validation metric of previous epoch
    \ENDFOR
  \end{algorithmic}
  \label{alg:SemiMol}
\end{algorithm}
\paragraph{Curriculum-based pseudo-labeling.} Meanwhile, the target molecular model $f$ receives discriminative information $\{p_i\}_{i=1}^{N+M}$ from the instructor model $g$ and uses it to efficiently select pseudo-labeled samples to backpropagate its gradient. Remarkably, the criterion leveraged to determine which subset of unlabeled samples $\mathcal{D}^{*}$ to be incorporated into the training in each round is vital to the success of pseudo-labeling. Various uncertainty metrics have been explored in the previous literature. For instance, several approaches choose instances with the highest-confidence~\cite{zhu2005semi} or retrieve the nearest samples in the feature space~\cite{shi2018transductive}. Afterward, some studies adopt techniques of fixed thresholds, tuning thresholds manually, or using percentile scores~\cite{cascante2021curriculum}.
Here, we propose a self-adaptive curriculum learning algorithm to enable pseudo-annotated samples to enter or leave the new training set. 

Specifically, we first initialize a percentage threshold $\gamma$ and use it to produce hard labels and constitute the hybrid set $\mathcal{D}'' = \mathcal{D}^{\star} \cup \left\{\left(x_i^{*},\hat{y}_i^{*}\right) | p_i^{*} \geq \gamma \right\}$ to train the target model $f$.

Then, as the training progresses, we adaptively adjust the confidence cut-off $\gamma$ based on the model's performance in the validation set. If the model is assessed to be more powerful (\emph{e.g.,} the evaluation metric like RMSE becomes smaller), we propose to lower the threshold as $\gamma = \gamma - \Delta \gamma$, where $\Delta \gamma$ is the stepping percent. This leads to a larger hybrid set $\mathcal{D}''$ and the loss of the target model $f$ can be written as:
\begin{equation}
    \mathcal{L}_f \left(\mathcal{D}''\right) = \sum_{(x''_i, y''_i) \in \mathcal{D}''} \mathcal{H}_f (f(x''_i), y''_i), 
\label{equ:loss}
\end{equation}
Our self-adaptive curriculum learning strategy discourages concept drift or confirmation bias since it can prevent erroneous annotations predicted by an undertrained network during the early stages of training to be accumulated~\cite{cascante2021curriculum}. The pseudo-code of our SemiMol is depicted in Algorithm~\ref{alg:SemiMol}.

\begin{figure*}[th]
    \centering
    \includegraphics[width=\textwidth]{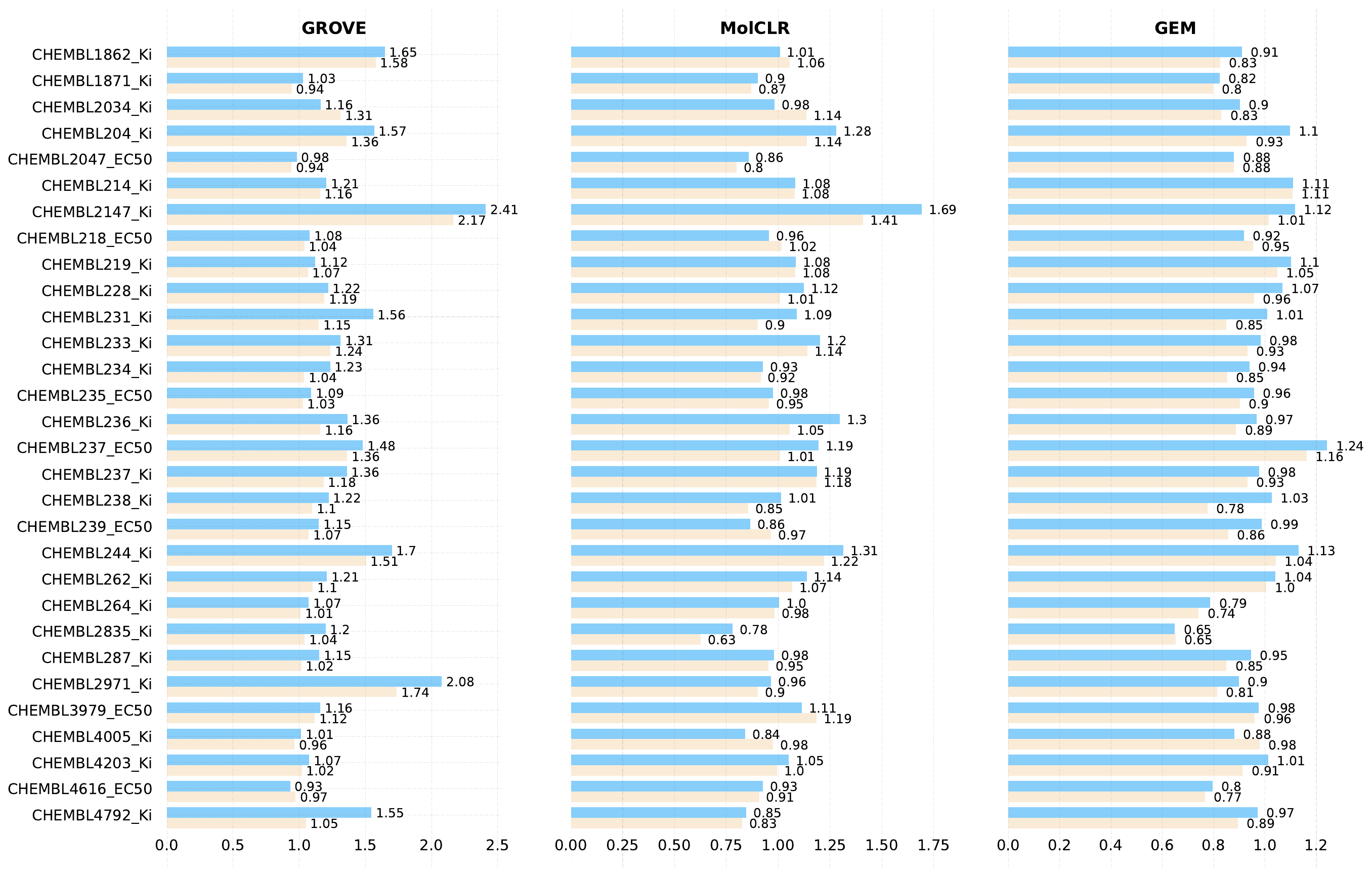}
    \caption{Comparison of different GNNs with and without pretraining on 30 activity cliff datasets. The light blue corresponds to the performance without pretraining, while the antique white represents the performance with pretraining. }
    \label{fig:pretrain}
\end{figure*}
\paragraph{Tips for implementing SemiMol.} Before executing SemiMol, it is natural to first obtain a well-trained molecular target model $f_0$ through regular supervised learning on the labeled dataset $\mathcal{D}^{\star}$ and then initialize an instructor model $g_0$ by discriminating pseudo-labels generated by that $f_0$. This is empirically proven to achieve higher training stability and robustness. Moreover, we choose to assign pseudo-labels every $k$ epochs, where a proper setting of $k$ is critical to the success of SemiMol. If pseudo-labels are updated too frequently, the training procedure tends to be volatile. 

\label{sec:performance}
\begin{figure*}[t]
    \centering
    \includegraphics[width=0.9\textwidth]{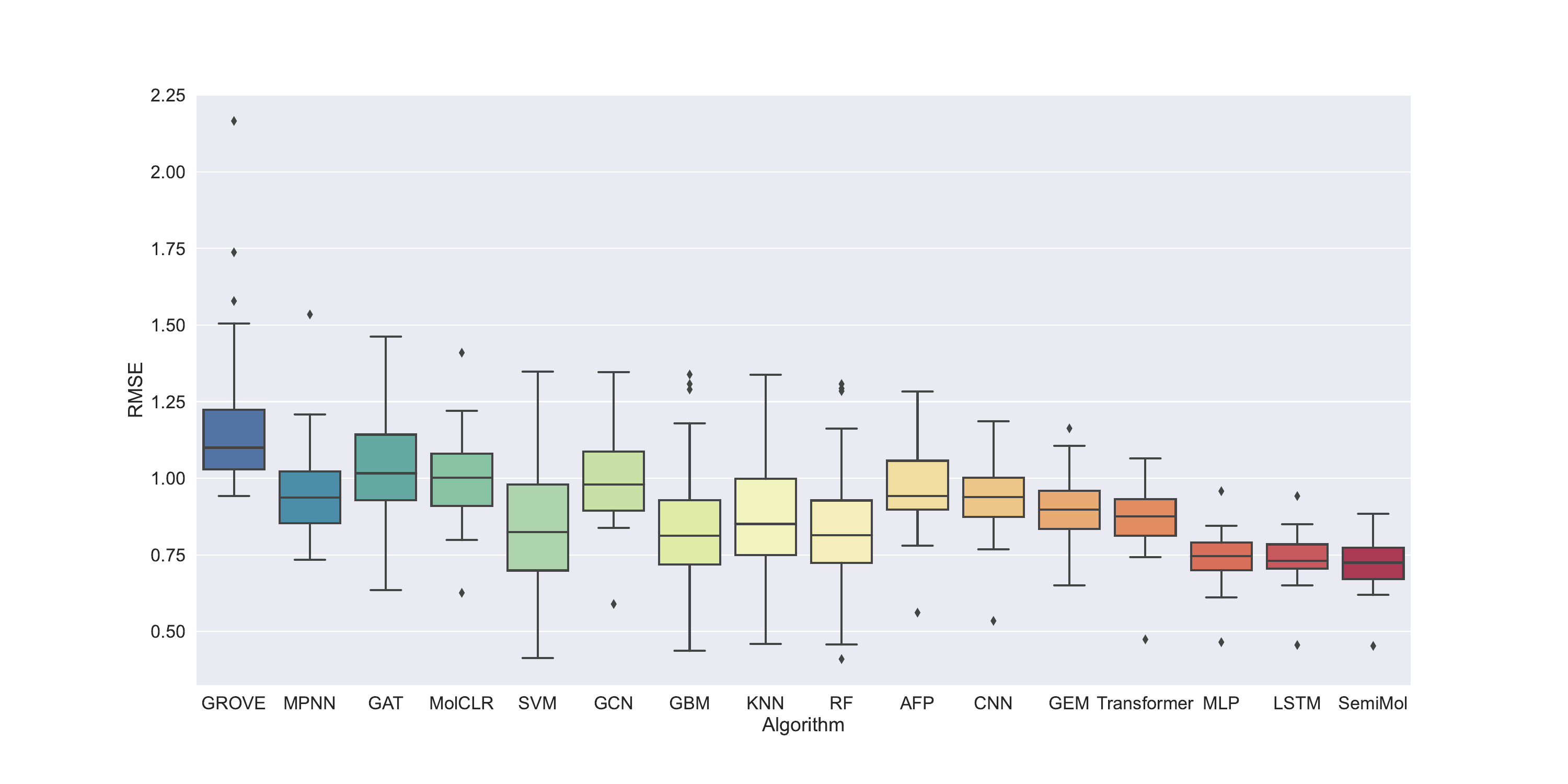}
    \caption{Performance of various ML models measured by the overall RMSE on 30 activity cliff datasets.}
    \label{fig:boxplot_1}
\end{figure*}
\section{Experiments}
\subsection{Data and Experimental Setups}
All our evaluations in this section are performed on datasets from MoleculeACE (Activity Cliff Estimation), which is an open-access benchmarking platform and available on Github~\url{https://github.com/molML/MoleculeACE}. It contains more than 35,000 molecules over 30
macromolecular targets, where each target corresponds to a dataset. In particular, 12 of 30 datasets in MoleculeCAE have no more than 1K molecules in the training set, indicating a standard low-data regime. In addition, we take Graph Isomorphic Network (GIN)~\cite{xu2018powerful} along with a Graph Multi-set Transformer (GMT)~\cite{baek2021accurate} to aggregate node features as the backbone. More experimental details and dataset statistics are elaborated on Appendix. 

\subsection{Does GNN Pretraining Help Mitigate the Activity Cliff Problem?}
\label{sec:pretrain}
Data quantity and quality are always regarded as one of the most pivot factors in the success of DL algorithms. However, assessing DL methods on activity cliffs is in a low-data scenario. It is worth noting that the past few years have witnessed the prevailing of self-supervised learning across multiple areas~\cite{devlin2018bert,brown2020language,dosovitskiy2020image}, 
and the graph research community in molecule modeling has been trying to replicate its success in natural language processing. Various works attempt to excavate information on large-scale unlabeled molecules~\cite{rong2020self,fang2022geometry,wang2022molecular,zhou2022uni}, and claim significant performance improvements~\cite{rong2020self,you2020does,zhu2021graph,zhou2022uni}. Nevertheless, the efficacy of those pretraining frameworks is only validated on datasets of non-activity cliff molecules and it remains unknown whether they are also powerful in alleviating the data scarcity trouble in activity cliff estimation. 

Here, we investigate several SOTA pretraining GNNs, containing GROVE~\cite{rong2020self}, MolCLR~\cite{wang2022molecular}, and GEM~\cite{fang2022geometry}, and examine their effectiveness in activity cliff datasets. To be explicit, GROVE adopts a Transformer-style~\cite{vaswani2017attention} architecture, significantly enlarging the representation ability and application scope of molecular representation learning schemes. MolCLR applies data augmentation to molecular graphs at both node and graph levels and uses a contrastive learning strategy to generalize GNNs to a more giant chemical space. GEM proposes a geometry-based GNN with dedicated geometry-level self-supervised learning techniques to capture molecular geometry knowledge. Figure~\ref{fig:pretrain} exhibits the empirical results. It can be found that pretraining generally benefits the activity cliff estimation problem in all three selected pretraining frameworks. It leads to average improvements of 8.57\%, 3.92\%, and 6.54\% for GROVE, MolCLR, and GEM on all 30 datasets. However, the improvements are not always convincing and can be negligible or even negative. For instance, GROVE performs worse with pretraining in two datasets, MolCLR becomes less competitive with pretraining in six datasets, and GEM is negatively influenced with pretraining in three datasets. 

This phenomenon is consistent with the conclusion in one recent study~\cite{sun2022does} that self-supervised graph pretraining does not always behave statistically more powerfully over non-pretraining methods in many settings. Besides, the local and global molecular 3D structures may not be the key to tackling our problem since GEM has incorporated them for self-supervised learning but shows little benefit. Finally, we compare the performance of different GNNs before and after self-supervised learning. GROVE reaches an average RMSE of 1.3107 and 1.1871 before and after pretraining. MolCLR attains an average RMSE of 1.0556 and 1.0071 before and after pretraining. GEM realizes an average RMSE of 0.9699 and 0.9035 before and after pretraining. The design of GNNs' architecture is a more foundational factor in achieving a better performance in our activity cliff problem. GEM without pretraining can easily outpace pretrained MolCLR and pretrained GROVE.

\paragraph{Why is GNN pretraining not Beneficial?} We fail to reproduce the gain of pretraining in conventional molecular datasets like MoleculeNet due to several potential reasons. Firstly, the target labels of most unsupervised pretraining are not appropriately aligned with or even contradictory to our activity cliff estimation task. For instance, GROVE~\cite{rong2020self} requires GNNs to decide whether a molecule contains a motif. However, these 85 sorts of pre-defined motifs might be deceptive in estimating the property of activity cliff molecules. Secondly, as suggested by~\cite{sun2022does}, pretraining is found to typically help when high-quality hand-crafted features are absent. In our experiments, we inject rich features into GNNs, including the additional node features such as hydrogen acceptor match, acidic match, and bond features such as ring information.  Last but not least, some pretraining methods, such as masked node label prediction, might be too easy to transfer enough knowledge for our activity cliff estimation task. 

\begin{table*}[t]
    \centering
    \resizebox{1.5\columnwidth}{!}{
    \begin{tabular}{l | ccccc}\toprule
    Datasets & CHEMBL1862\_Ki & CHEMBL1871\_Ki & CHEMBL2034\_Ki & CHEMBL204\_Ki & CHEMBL2047\_EC50  \\  \midrule
     \textbf{GIN} &  ${1.019}$ & ${0.854}$ & ${0.951}$ & ${1.179}$ & ${0.895}$  \\
     + $\pi$-model & ${0.984}$ & ${0.793}$ & ${0.902}$ & ${1.060}$ & ${0.880}$ \\ 
     + Semi-GAN & $0.955$ & ${0.781}$ & ${0.834}$ & ${0.927}$ & ${0.854}$ \\ 
     + SemiMol & $\mathbf{0.714}$ & $\mathbf{0.671}$ & $\mathbf{0.703}$ & $\mathbf{0.801}$ & $\mathbf{0.629}$ \\ \bottomrule 
    \end{tabular}}
    \caption{Performance of different semi-supervised learning algorithms on 5 datasets of MoleculeACE. }
    \label{tab:compare_ssl}
\end{table*}
\begin{table*}[t]
    \centering
    \resizebox{1.5\columnwidth}{!}{
    \begin{tabular}{l | ccccc}\toprule
     Datasets & CHEMBL214\_Ki & CHEMBL2147\_Ki & CHEMBL218\_EC50  & CHEMBL219\_Ki & CHEMBL228\_Ki  \\ \midrule
     \textbf{SemiMol} &  \\
     + Fixed Threshold & ${0.788}$ & $0.704$ & ${0.781}$ & ${0.914}$ & ${0.821}$ \\ 
     + Percentile Scores & $0.739$ & ${0.683}$ & ${0.744}$ & ${0.850}$ & ${0.779}$ \\ 
     + Self-adaptive Paces & $\mathbf{0.726}$ & $\mathbf{0.672}$ & $\mathbf{0.723}$ & $\mathbf{0.835}$ & $\mathbf{0.758}$ \\ \bottomrule
    \end{tabular}}
    \caption{Comparison with different curriculum learning methods on 5 datasets of MoleculeACE. }
    \label{tab:ablation}
\end{table*}

\subsection{Does SemiMol Improve the Activity Cliff Estimation?}
\paragraph{Overall performance.} We compare SemiMol with diverse ML and DL algorithms on MoleculeACE and document the results of activity cliff molecules and all molecules in Figure~\ref{fig:boxplot_1} and Appendix, separately. It can be observed that our SemiMol outperforms other baselines with the lowest RMSE on 30 activity cliff datasets. More essentially, we discover an average improvement of 26.53\% brought by SemiMol, which is significantly larger than the benefits of SOTA pretraining approaches. This phenomenon declares that GNNs empowered by SSL can be a more influential toolkit than existing algorithms to overcome the obstacle of activity cliffs and accelerate the process of new drug discovery.  

Among all approaches, non-pretrained GNNs, including MPNN~\cite{gilmer2017neural}, GAT~\cite{velivckovic2017graph}, GCN~\cite{kipf2016semi}, and AFP~\cite{baek2021accurate}, generally perform worse than other methods. Meanwhile, the performance of pretrained GNNs varies dramatically. For instance, GEM outweighs GROVE by a large margin. However, as analyzed in Section~\ref{sec:pretrain}, the benefit of pretraining is proven limited. This difference is more related to the capability of the backbone GNNs than to the pretraining strategies.  Additionally, sequence-based models such as LSTM~\cite{hochreiter1997long} and Transformer~\cite{vaswani2017attention} are conspicuous for their outstanding performance. They are empirically a better choice for addressing the activity cliff problem than graph-based and descriptor-based models if no additional unlabeled data are provided in the fine-tuning stage. 

\paragraph{Comparison with other semi-supervised methods.} Diverse SSL algorithms have been invented to employ the abundant unlabeled resources in a task-specific manner. Approximately, they can be divided into three categories: consistency regularization, proxy-label methods, and generative models. Here, we pick up one baseline instance from each kind and investigate their improvements for molecular property prediction. Notably, random node dropping and edge perturbation are biochemically improper, and data augmentation may alter the semantics of molecular graphs~\cite{sun2021mocl}. Thus, we employ the $\pi$-model~\cite{laine2016temporal} for comparison, which applies different stochastic transformations (\emph{i.e.}, dropout) to the networks instead of the input graphs. Moreover, we select Semi-GAN~\cite{odena2016semi}, which introduces a discriminator to classify whether the input is labeled or not. As mentioned previously, UPS~\cite{rizve2021defense} leverages the prediction uncertainty to guide the pseudo-label selection procedure and is therefore inapplicable in our situation. We select five datasets from MoleculeACE for comparison and use GIN as the backbone architecture. As shown in Table~\ref{tab:compare_ssl}, SemiMol outperforms all other semi-supervised methods in overall RMSE. In addition to that, SemiMol overcomes the drawbacks of UPS and can be used for regression tasks. All the above evidence clarifies that SemiMol is a cutting-edge proxy labeling algorithm in comparison to existing SSL mechanisms with stronger generalization and broader applications. 

\paragraph{Ablation study.} We also implement an extra experiment to illustrate the effectiveness of our self-adaptive curriculum learning mechanism and report the results in Table~\ref{tab:ablation}. It can be found that our self-adaptive schedule achieves stronger performance in overall RMSE than a fixed threshold and percentile score-based curriculum learning paradigm~\cite{cascante2021curriculum}.

\paragraph{Performance for classification tasks.} Notably, datasets in MoleculeACE are all regression tasks, and one major advantage of our SemiMol over other proxy-labeling algorithms is that SemiMol can deal with continuous property space. In spite of that, SemiMol can also be employed for classification tasks. In order to better demonstrate its superiority, we use a classification dataset, CYP3A4~\cite{rao2022quantitative}, and examine its performance. CYP3A4 consists of the property cliffs in the Cytochrome P450 3A4 inhibitions experimentally measured by~\cite{veith2009comprehensive}, including 3,626 active inhibitors/substrates and 5,496 inactive compounds. We report the corresponding performance of different approaches in Table~\ref{tab:CYP3A4}, where the best performance is in bold and the second best is underlined. It can be observed that SemiMol outperforms all semi-supervised baselines, which rely on MPNN as the backbone architecture. 
\begin{table}[ht]
    \centering
    \resizebox{0.7\columnwidth}{!}{
    \begin{tabular}{c cc  }\toprule
         \multirow{2}{*}{Models} & \multicolumn{2}{c}{\textbf{CYP3A4}}  \\ 
         & ROC-AUC & Cliff ROC-AUC \\ \midrule
        GCN & 0.7659 &  0.6801\\
        GAT & 0.7734 &  0.6828 \\
        MPNN & {0.7811} &  {0.6867} \\ \midrule 
        $\pi$-model & 0.7842 & 0.6903 \\
        Semi-GAN & 0.8015 & 0.7122 \\
        UPS & \underline{0.8438} &  \underline{0.7546} \\ \midrule
        SemiMol & \textbf{0.8568} & \textbf{0.7703} \\ \bottomrule 
    \end{tabular}}
    \caption{Performance in the classification dataset, CYP3A4. }
    \label{tab:CYP3A4}
\end{table}

\section{Conclusion}
The past decade has witnessed the increasing employment of machine learning techniques for drug discovery, but the topic of activity cliffs (AC) receives little interest from the scientific community. In this work, we identify the best practice to enhance models’ predictivity in the presence of activity cliffs and propose an effective semi-supervised learning (SSL) method named SemiMol. It introduces an instructor model to understand the confidence of proxy-label and exploits a self-adaptive curriculum training paradigm to move the target model toward hard samples more efficiently. We demonstrate that graph-based DL models can hold up against simpler ML algorithms for drug discovery even in low-data scenarios. We envision more efficient and better-suited SSL methods to propel the frontier of AC identification. 

\bibliographystyle{named}
\bibliography{cite}
\newpage
\onecolumn
\section*{Experimental Setting}
\label{app:exp}
All experiments were run on two A100-GPUs. We use Pytorch Geometric package (PyG)~\cite{fey2019fast} v.2.2.0 to build all GNNs. An Adam optimizer is used to propagate the gradient. We try a ReduceLROnPlateau scheduler to adjust the learning rate with a factor of 0.6 and a patience of 10, but find a constant learning rate also performs well. Hyperparameters to search in running the experiments of SemiMol are listed in Table~\ref{tab:hyper}.
\begin{table}[ht]
    \caption{Hyperparameters setup for SemiMol.}
    \centering
    \begin{tabular}{lll}\toprule
    Hyperparameters Search Space & Symbol & Value \\ \midrule
    \textbf{Training Setups} \\ 
    Learning rate & - & [1e-4, 5e-5, 1e-6, 5e-6]\\ 
    Batch size & - & [32, 64, 128]\\ 
    Epochs & - & [100, 300]\\ 
    Dropout rate & - &  [0.2, 0.4] \\ 
    \textbf{SemiMol Setups} \\ 
    Loss weight & $\lambda$ &  [0.1, 1.0, 10.0, 100.0]\\ 
    Starting Confidence Threshold & $\gamma$ & [0.8, 0.9, 0.95] \\ 
    Stepping Confidence Percent & $\Delta \gamma$ & [0.05, 0.1, 0.2] \\ 
    Pseudo-label Update Frequency & $k$ &  [5, 10, 15, 20]\\ 
    \textbf{Neural Network Setups} \\ 
    Number of MPNN layers & - & [2, 3, 4, 5, 6]\\
    The hidden dimension of node representations & - & [32, 64, 128, 256, 512]\\
    The hidden dimension of edge representations & - & [64, 128, 256]\\
    Number of heads in GMT & - & [4, 8 ,12]\\
    Hidden dimension in GMT & - & [64, 128, 256, 512] \\
    Number of fully-connected layers & - & [1, 2, 3]\\ \bottomrule 
    \end{tabular}
    \label{tab:hyper}
\end{table}

\section*{Baselines}
Here we provide detailed descriptions of some key baselines in the ACE problem. 

\subsection*{Descriptor-based Models}
\textbf{KNN, SVM, GBM, RF, and MLP.} Each algorithm was combined with four types of molecular descriptors, \emph{i.e.}, human-engineered numerical features designed to capture predetermined chemical information. We explored molecular descriptors with several levels of complexity: (1) extended connectivity fingerprints (ECFPs), encoding atom-centered radial substructures in the form of a binary array; (2) Molecular ACCess System (MACCS) keys, which encode the presence of predefined substructures in a binary array; (3) weighted holistic invariant molecular (WHIM) descriptors, capturing three-dimensional molecular size, shape, symmetry, and atom distribution; and (4) eleven physicochemical properties relevant for drug-likeness.

\subsection*{Graph-based Models}
\textbf{GCN, GAT, MPNN, and AFP} are all implemented using the Pytorch Geometric package. In GCN, GAT, and MPNN, the global pooling is realized using a graph multiset Transformer. 

\subsection*{SMILES-based Models}
All SMILES strings are encoded as one-hot vectors. SMILES strings longer than 200 characters are truncated (0.71\% on average). Noncanonical SMILES strings are generated using RDKit~\cite{landrum2013rdkit}. \textbf{LSTM}~\cite{hochreiter1997long} is a type of recurrent neural network (RNN), which can learn from string sequences by keeping track of long-range dependencies. In our study, LSTM models were pretrained on SMILES obtained by merging all training sets with no repetitions (36K molecules) using next-character prediction before applying transfer learning for bioactivity prediction. \textbf{Transformer}~\cite{vaswani2017attention} processes the whole sequence at once in a graphlike manner using positional embedding to capture positional information.  Transformers implement the self-attention mechanism to learn which portions of the sequence are more relevant for a given task. The ChemBERTa architecture, which was pretrained on 10M compounds, was used in combination with transfer learning for bioactivity prediction. \textbf{1D CNN} utilizes a single 1D convolutional layer with a kernel size equal to 1 followed by a fully connected layer. 

\section*{Data}
\label{app:data}
\subsection*{Unlabeled Data}
 We use the ZINC15~\cite{sterling2015zinc} database to collect unlabeled molecular data, which can be downloaded from DeepChem~\cite{Ramsundar-et-al-2019}. There are four different data sizes supported by ZINC15: 250K, 1M, 10M, and 270M. We use the 250K version as the unlabeled dataset for simplicity. Those unlabeled SMILES are then converted by RDKit~\cite{landrum2013rdkit} into 2D graphs. We expect future work to enrich the unlabeled corpus by leveraging other resources such as ChemBL~\cite{gaulton2012chembl}, Chembridge, and Chemdiv.

\subsection*{Activity Cliff Data}
MoleculeACE is a tool for evaluating the predictive performance on activity cliff compounds of ML models. It collects and curates bioactivity data on 30 macromolecular targets. The curated collection contains a total of 48.7K molecules, of which 35.6K are unique, and mimics typical drug discovery datasets as it includes several target families and spans different training scenarios. For each macromolecular target, activity cliffs are identified by considering pairwise structural similarities and differences in potency. Three distinct approaches containing substructure similarity, scaffold similarity, and similarity of SMILES strings are used to quantify molecular similarities between any pairs of molecules.

Since MoleculeACE is a relatively new benchmark with no prior work thoroughly calculating its statistics, we compute the key stats in Table~\ref{tab:data} for readers to have a comprehensive understanding of its molecular distributions. All datasets in MoleculeACE use the split as suggested by~\cite{van2022exposing}. The data statistics and a brief description of those datasets are reported in Table~\ref{tab:data}. We refer the readers to~\cite{van2022exposing} for more details regarding the data curation and detection of activity cliffs.  
\begin{table}[ht]
    \caption{Description of 30 regression datasets in MoleculeACE, including the number of molecules, the maximum, the minimum, the mean, and the standard deviation of the ground truth properties. }
    \centering
    \begin{tabular}{l c c c c c c} \toprule
    &  \# of Compounds & Min & Max & Mean & Std \\ \midrule
    CHEMBL4203\_Ki & 731 & -4.90 & 0.50 & -2.50 & 1.02 \\ 
    CHEMBL2034\_Ki & 750 & -4.27 & 1.00 & -1.20 & 1.09 \\ 
    CHEMBL233\_Ki & 3,142 & -4.80 & 2.00 & -1.68 & 1.29 \\ 
    CHEMBL4616\_EC50 & 682 & -4.15 & 1.00 & -1.20 & 0.93 \\ 
    CHEMBL287\_Ki & 1,328 & -4.67 & 1.44 & -1.52 & 1.09 \\ 
    CHEMBL218\_EC50 & 1,031 & -4.99 & 1.52 & -2.25 & 1.05 \\ 
    CHEMBL264\_Ki & 2,862 & -4.93 & 1.60 & -1.25 & 1.10 \\ 
    CHEMBL219\_Ki & 1,859 & -4.95 & 1.74 & -1.77 & 1.07 \\ 
    CHEMBL2835\_Ki & 615 & -3.66 & 1.00 & -0.33 & 0.95 \\ 
    CHEMBL2147\_Ki & 1,456 & -5.00 & 2.00 & -1.14 & 1.96 \\ 
    CHEMBL231\_Ki & 973 & -5.00 & 1.31 & -2.07 & 1.26 \\ 
    CHEMBL3979\_EC50 & 1,125 & -4.79 & 1.22 & -2.27 & 1.15 \\ 
    CHEMBL237\_EC50 & 955 & -4.70 & 1.96 & -1.56 & 1.39 \\ 
    CHEMBL244\_Ki & 3,097 & -5.00 & 2.00 & -1.86 & 1.64 \\ 
    CHEMBL4792\_Ki & 1,471 & -4.25 & 1.15 & -2.07 & 1.11 \\ 
    CHEMBL1871\_Ki & 659 & -4.75 & 0.60 & -1.95 & 1.03 \\ 
    CHEMBL237\_Ki & 2,602 & -4.91 & 1.85 & -1.66 & 1.33 \\ 
    CHEMBL262\_Ki & 856 & -5.00 & 1.05 & -2.49 & 1.06 \\ 
    CHEMBL2047\_EC50 & 631 & -4.96 & 0.52 & -2.77 & 0.98 \\ 
    CHEMBL239\_EC50 & 1,721 & -4.95 & 1.59 & -2.65 & 1.09 \\ 
    CHEMBL2971\_Ki & 976 & -5.00 & 1.22 & -1.14 & 1.45 \\ 
    CHEMBL204\_Ki & 2,754 & -6.39 & 2.00 & -2.10 & 1.51 \\ 
    CHEMBL214\_Ki & 3,317 & -4.80 & 1.85 & -1.66 & 1.14 \\ 
    CHEMBL1862\_Ki & 794 & -5.00 & 1.73 & -1.41 & 1.51 \\ 
    CHEMBL234\_Ki & 3,657 & -4.86 & 1.70 & -1.54 & 1.18 \\ 
    CHEMBL238\_Ki & 1,052 & -4.97 & 0.35 & -2.34 & 1.09 \\ 
    CHEMBL235\_EC50 & 2,349 & -5.00 & 1.74 & -2.57 & 1.09 \\ 
    CHEMBL4005\_Ki & 960 & -4.35 & 1.52 & -1.31 & 1.07 \\ 
    CHEMBL236\_Ki & 2,598 & -5.00 & 2.00 & -2.01 & 1.35 \\ 
    CHEMBL228\_Ki & 1,704 & -4.86 & 1.89 & -1.67 & 1.21 \\ \bottomrule
    \end{tabular}
    \label{tab:data}
\end{table}

In CYP3A4, the active compounds were compared with inactive ones through MMPA46 in the RDKit package~\cite{landrum2013rdkit}. This led to 106 molecular pairs and corresponding substructures involving 46 active and 51 inactive compounds. During the GNN model training, all these involved compounds involved in found pairs are put into the test set. CYP3A4 can be attained in the Github of Mol-XAI~\cite{rao2022quantitative}~\url{https://github.com/biomed-AI/MolRep} and the data is stored in~\url{https://drive.google.com/drive/folders/1Lbb5KBat4Au_1yMHKyYnAxf6xVFNWtkW}. 

To make a thorough comparison, we pick up random forest (RF), gradient boosting (GBM), support vector regression (SVM), K-nearest neighbor (KNN), message passing neural network
(MPNN)~\cite{gilmer2017neural}, graph convolutional network (GCN)~\cite{kipf2016semi}, graph attention network (GAT)~\cite{velivckovic2017graph}, attentive fingerprint (AFP)~\cite{xiong2019pushing}, long short-term memory networks (LSTM)~\cite{hochreiter1997long}, 1D convolutional neural networks (CNN)~\cite{kimber2021maxsmi}, and Transformers~\cite{vaswani2017attention} as baselines. We run a Bayesian optimization with a Gaussian process to optimize the hyperparameters as suggested in MoleculeACE~\cite{van2022exposing}. The details of those architectures are listed as follows: 
\begin{itemize}
    \item KNN, SVM, GBM, and RF regression models were implemented using sklearn~\cite{pedregosa2011scikit}. For KNN, we optimize the number of neighbors $k$ from the candidate $k=[3,5,11,21]$. For SVM,  we optimize the kernel coefficient $\gamma$ and regularization parameter $C$ from the candidate $\gamma=\left[1 \times 10^{-6}, 1 \times 10^{-5}, 1 \times 10^{-4}, 1 \times 10^{-3}, 1\right.$ $\times 10^{-2}$, or $\left.1 \times 10^{-1}\right]$ and $C=[1,10,100,1000,10,000]$. For GBM, we optimize the number of boosting stages $n_{\mathrm{b}}$ and the maximal model depth $m_{\mathrm{d}}$ from the candidate values $n_{\mathrm{b}}=[100,200,400]$ and $m_{\mathrm{d}}=[5$, $6,7]$. For RF, we select the number of decision trees $t$ from the candidate list $t=[100,250$, $500,1000]$.
    \item For sequence-based methods, SMILES strings were encoded as one-hot vectors. SMILES strings longer than 200 characters were truncated. LSTM model has 4 layers with hidden units of 1024, 1024, 256, and 256.  Transformer models and the corresponding SMILES tokenization were based on the ChemBERTa~\cite{baek2021accurate}.  
    \item For graph-based baselines, we mainly search for hidden atom features from $[32,64,128, 256,512]$, the number of convolutional layers from $[1,2,3,4,5]$, hidden multiset Transformer nodes from $[64,128,256,512]$ and hidden predictor features from $[128,256,512]$. For AFP, we tune the timestep from $[1,2,3,4,5]$.
\end{itemize}

We also select three SOTA pretraining GNNs to examine the efficacy of pretraining on activity cliffs. GROVE~\cite{rong2020self} is downloaded from its official Github~\url{https://github.com/tencent-ailab/grover}. 
To be explicit, we use the base pretrained model for GROVE. MolCLR is downloaded from its official Github~\url{https://github.com/yuyangw/MolCLR}
GEM is downloaded from its official Github~\url{https://github.com/PaddlePaddle/PaddleHelix/tree/dev/apps/pretrained_compound/ChemRL/GEM}. Both GROVE and MolCLR are implemented in PyG, while GEM is implemented in its self-built PaddlePaddle platform.~\cite{ma2019paddlepaddle}.

\section*{Activity Cliffs on Proteins}
In the main text, we mainly discuss the divergent performance of machine learning algorithms between non-activity cliff small molecules and activity cliff small molecules. However, this phenomenon can also exist for macromolecules, such as proteins. For instance, some prior studies~\cite{shortle2009one} discovered that a single amino acid substitution could completely change the fold of a protein. And the structural change produced by this one mutation cannot be dismissed as a semantic issue over what constitutes a different fold. To be explicit, one conformation consists of a three-helix bundle, whereas the alternate form has a four-strand $\beta$-sheet with a single $\alpha$-helix. This one amino acid can cause a big structural difference even though other segments in the sequence are the same. This ``similar sequence, different property (\emph{i.e.}, structures)'' can be a big challenge for tasks like protein structure prediction, which can be seen as a different version of activity cliffs in macromolecules. We envision more future work that can explore the representation behavior of machine learning algorithms in modeling this type of macromolecules and examine whether they suffer from the same issue as representing activity cliff small molecules.


%

\end{document}